\documentclass[journal]{IEEEtran}
\IEEEoverridecommandlockouts
\usepackage{cite}
\usepackage{amsmath,amssymb,amsfonts}
\usepackage{algorithmic}
\usepackage{graphicx}
\usepackage{textcomp}
\usepackage{booktabs}

\usepackage{bm}

\def\BibTeX{{\rm B\kern-.05em{\sc i\kern-.025em b}\kern-.08em
    T\kern-.1667em\lower.7ex\hbox{E}\kern-.125emX}}
\begin{document}

\title{Temporal Convolutional Autoencoder for Interference Mitigation in FMCW Radar Altimeters

\thanks{Portions of this work were presented at IEEE MILCOM 2024, Washington DC, USA \cite{brown2024aircraft}. Corresponding author: Charles E. Thornton (e-mail: cthorn14@vt.edu).}

\author{\IEEEauthorblockN{Charles E. Thornton\IEEEauthorrefmark{1}, Jamie Sloop\IEEEauthorrefmark{1}, Samuel Brown\IEEEauthorrefmark{1},\\ Aaron Orndorff\IEEEauthorrefmark{1}, William C. Headley\IEEEauthorrefmark{1}, Stephen Young\IEEEauthorrefmark{2}}\\
\IEEEauthorrefmark{1}Virginia Tech National Security Institute,
Blacksburg, VA, USA \\
\IEEEauthorrefmark{2}The Boeing Company, Tukwila, WA, USA}}

\maketitle
\thispagestyle{plain}
\pagestyle{plain}

\begin{abstract}
Reliable altitude estimation with frequency-modulated continuous wave (FMCW) radar altimeters is increasingly a challenge due to in-band interference from modern communication systems. In this paper, we present a temporal convolutional autoencoder (TCAE) that directly processes in-phase and quadrature (IQ) samples to suppress structured interference while preserving signal phase and frequency content for range estimation. The model is trained and initially evaluated within a full radar altimeter simulation chain, then further validated via over-the-air (OTA) experiments using a universal software radio peripheral (USRP)-based testbed. Results show that the TCAE reduces altitude estimation error by more than 85\% compared to least mean squares (LMS) adaptive filtering under severe interference conditions, including low signal-to-interference-plus-noise ratio (SINR) and full temporal overlap between interfering and radar signals. Unlike conventional methods, the TCAE maintains phase fidelity and beat structure, enabling accurate range estimation even when interferers occupy more than one-quarter of the radar bandwidth. The implemented TCAE performs mitigation directly on fixed-length IQ windows using a single feed-forward pass and was integrated into the MATLAB/ONNX-based evaluation chain used for both simulation and OTA testing. These findings demonstrate that learned IQ-domain interference mitigation can enhance radar-altimeter resilience under a range of tested interference conditions.
\end{abstract}


\section{Introduction and Motivation}
Precise and reliable altitude measurements are essential in a number of airborne applications. Radar altimeters are widely employed to obtain such measurements for objectives such as commercial aviation, the operation of unmanned aerial systems, and military missions at low altitudes \cite{quartly2021overview}. Due to the sensitive nature of these and related applications, it is crucial to understand all potential sources of system failure and performance degradation. 

A particularly notable threat to reliable radar altimeter operation is radio-frequency (RF) interference from neighboring systems, which could be either intentional or incidental \cite{singh2017method}. Commercial altimeters commonly operate in the $4.2$-$4.4$ GHz band, which has created concern among regulators given the increased proliferation of communication systems in the $1$-$6$ GHz spectrum. For example, in January 2023, the Federal Aviation Administration (FAA) issued a deadline for all commercial airlines to upgrade altimeters to ensure safe operation in the vicinity of 5G C-Band signals \cite{FAA2024_5G}. By September 2023, the entire US airline fleet had updated their altimeter equipment to ensure that the risk of 5G interference would be mitigated through 2027. However, the possibility of interference from other sources, the use of radar altimeters in wider applications, and the growing congestion of the sub 6 GHz spectrum all provide pressing demand for robust interference mitigation strategies in future generations of radar altimeters.

Interference mitigation in radar systems is an important and broad area of research \cite{zheng2019radar}. Conventional approaches for interference suppression include time- and frequency-domain methods such as adaptive filtering, spectral masking, and wavelet-based methods. In the spatial domain, antenna-based techniques such as phased array beamforming and null steering may be applied. Another approach is to detect and cancel interference by monitoring changes in amplitude, frequency, or phase. While effective in specific scenarios, such approaches often rely on specific assumptions, frequently require careful parameter tuning, and risk degrading the radar return when interference overlaps with the target signal.

In recent years, denoising autoencoders (DAEs) have emerged as a compelling approach for interference mitigation \cite{oyedare2022interference}. DAEs learn a compact, low-dimensional representation of the desired signal during a training process, enabling them to adapt to interference patterns that are too complex for traditional filters. Additionally, once the model is trained, interference mitigation can be performed by a single forward pass through the model, which can be faster than iterative methods such as adaptive filtering. However, most existing studies for radar interference mitigation with DAEs apply autoencoders at intermediate stages, rather than directly on in-phase and quadrature (IQ) data. Additionally, system-level validation, particularly with over-the-air (OTA) measurements, remains rare. These limitations motivate the present study. We demonstrate a novel IQ-domain temporal convolutional autoencoder (TCAE) for frequency-modulated continuous wave (FMCW) radar altimeters, validated both in full-chain simulation and OTA experiments. While the present study focuses on FMCW altimeters, the IQ-domain processing approach is not inherently waveform-specific, and extension to other radar types represents a natural direction for future investigation.

However, there are several challenges associated with the use of DAEs for interference mitigation. First, the training data is required to be diverse and representative of dynamic real-world conditions. Second, the training phase can be computationally intensive and difficult to scale to wide-bandwidth signals. Third, DAEs often lack interpretability, which could create concerns for sensitive applications such as aviation. Thus, while DAEs show promise, prior work has rarely demonstrated IQ-domain interference mitigation in radar altimeters with end-to-end system validation, especially in over-the-air conditions. This motivates the present study.

\subsection*{Contributions}
While prior radar altimeter interference mitigation studies have focused on modeling, standards, or post-processed signals, few have demonstrated interference suppression directly on IQ data with end-to-end evaluation and over-the-air (OTA) validation. This work addresses that gap. The main contributions are:
\begin{enumerate}
    \item \emph{IQ-domain denoising without pre-processing}: Unlike deep learning methods that operate on beat (intermediate frequency, IF) signals or range-Doppler maps, the proposed TCAE takes IQ samples directly as input, prior to dechirping. This preserves the phase and frequency content required for accurate range estimation and eliminates the need for interference detection or signal transformation as a prerequisite. 
    \item \emph{Full-chain integration with altitude-referenced evaluation}: The TCAE is embedded directly into a complete FMCW radar altimeter signal chain, including realistic terrain and flight trajectory data. Performance is evaluated by altitude estimation root mean square error (RMSE) over a simulated landing scenario, rather than signal reconstruction error alone. This end-to-end evaluation framework provides system-level validation that goes beyond the signal-domain metrics typically reported in prior autoencoder-based interference mitigation studies.
    \item \emph{Demonstrated simulation-to-hardware generalization via OTA experiments}: Using a universal software radio peripheral (USRP)-based testbed, we demonstrate that the TCAE trained exclusively on synthetic data suppresses high-power in-band interference while preserving altitude estimation accuracy on hardware captures. This confirms that the learned representations are not artifacts of the simulation environment and establishes a proof of concept for deployment without hardware-specific retraining.
\end{enumerate}

\section{Prior Art}
Interference mitigation techniques for radar have been extensively studied. However, the field remains an active area of research and development, largely due to growing spectrum congestion and the stringent performance demands imposed by emerging high-resolution radar applications. A broad survey of recent techniques is provided in \cite{zheng2019radar,Griffiths2015}. 

For radar altimeters in particular, prior work spans physics-based interference modeling, regulatory studies, and machine learning approaches. Simulation frameworks such as \cite{remcom} emphasize electromagnetic propagation effects, including terrain and environmental factors, and use ray-tracing and full-wave solvers to evaluate altimeter performance under interference. Regulatory analyses \cite{rtca} define performance and certification requirements for operation in shared spectrum environments, ensuring resilience against coexisting communication systems. 

Building on these foundations, recent studies have explored data-driven methods. Amaireh \emph{et al.} \cite{amaireh2024machine} propose machine learning (ML)-based approaches for altimeter signal processing, while Rock \emph{et al.} \cite{rock2024} present a convolutional neural network (CNN) autoencoder for suppressing adjacent-band interference such as 5G signals. Together, these efforts highlight both the established value of physics-based modeling and the growing promise of deep learning, but they also reveal a gap: few methods operate directly on raw IQ signals and validate performance in both simulation and OTA, as pursued here.

Autoencoder architectures have been used previously for RF denoising and feature extraction. Kokalj-Filipovic \emph{et al.} \cite{Kokalj2019} applied denoising autoencoders to RF signal classification, demonstrating compact feature learning. Fuchs \emph{et al.} \cite{Fuchs2020} used a convolutional autoencoder for radar interference mitigation, but restricted processing to range-Doppler images, potentially losing time-domain information. Chen \emph{et al.} \cite{Chen2021} proposed a CNN-based autoencoder with gated convolutions for automotive radar, operating on the beat signal after interference detection. In contrast, our approach applies the autoencoder directly to the raw time-domain signal, removing the need for pre-processing or interference detection. 

Temporal Convolutional Networks (TCNs) extend these ideas by explicitly modeling temporal dependencies. Compared with recurrent models, TCNs use convolutional operations with dilation to capture long-range temporal structure efficiently \cite{Lea2016}. They have been applied to domains such as speech recognition, anomaly detection, and traffic forecasting \cite{Chauhan2023}, and Park \emph{et al.} \cite{Park2022} showed that combining TCNs with autoencoders can overcome bottlenecks of recurrent models for time-series data. These findings suggest strong potential for TCAEs in radar applications with structured temporal correlations.

Finally, traditional radar interference mitigation remains a key benchmark. Griffiths \emph{et al.} \cite{Griffiths2015} provide a comprehensive review of spectrum engineering and coexistence strategies, while Uysal and Sanka \cite{Uysal2018} demonstrate suppression for automotive radar using time-frequency methods. These classical approaches remain relevant for evaluating deep learning-based interference suppression and emphasize the continued need for robust and adaptable techniques.

%Interference mitigation: \cite{Griffiths2015,Uysal2018}

\section{Radar Altimeter System Model}
We consider a radar altimeter operating at center frequency $f_{c}$ in the $4.2$-$4.4$ GHz band with a fixed instantaneous bandwidth $B$. The altimeter produces measurements over discrete time steps $m = 1,2,\ldots,n$.

\subsection{Signal Model}
To describe the triangular FMCW waveform, we define several timing variables.  
Let $T_{\text{sw}}$ be the sweep period and $T_h = T_{\text{sw}}/2$ the half-sweep duration.  
The local sweep time is
\[
u = t - \Big\lfloor \tfrac{t}{T_{\text{sw}}}\Big\rfloor T_{\text{sw}}, 
\qquad u \in [0,\,T_{\text{sw}}),
\]
with sweep sign
\[
\sigma(u) = \begin{cases}
+1, & 0 \le u < T_h, \\[6pt]
-1, & T_h \le u < T_{\text{sw}},
\end{cases}
\]
and folded half-sweep time
\[
\tilde u = \min(u,\, T_{\text{sw}} - u), \qquad \tilde u \in [0,\,T_h].
\]

The transmit FMCW signal is
\begin{equation}
s_T(t) = A_{\text{tx}} \cos\!\left( 2\pi \left[ f_c t + \tfrac{k}{2}\,\sigma(u)\,\tilde u^{\,2} \right] \right),
\label{eq:tx}
\end{equation}
where $A_{\text{tx}}$ is the transmit amplitude and $k = B/T_h$ is the chirp slope in Hz/s.

The received signal is
\begin{multline}
S_R(t) = A_{\text{rx}} 
\cos\!\Bigg( 2\pi \Big[ f_c (t-\tau) 
+ \tfrac{k}{2}\,\sigma(u-\tau)\,\tilde u^{\,2} \\
\hspace{3.5em} +\, f_D t \Big] + \phi_0 \Bigg) \\
+\, s_{\text{intf}}(t) + s_{\text{clutter}}(t) + n(t),
\label{eq:rx}
\end{multline}
where $\tau = 2R/c$ is the two-way propagation delay, $f_D$ is the narrowband Doppler frequency shift which is modeled as an additive phase term, and $\phi_0$ is the initial phase. The terms $s_{\text{intf}}(t)$, $s_{\text{clutter}}(t)$, and $n(t)$ denote interference, clutter, and noise, respectively.

After dechirping, the beat frequencies for the upsweep and downsweep are
\begin{equation}
f_{b,\uparrow} = k\tau + f_D, 
\qquad 
f_{b,\downarrow} = -k\tau + f_D.
\label{eq:beat}
\end{equation}

Averaging the upsweep and downsweep estimates cancels most Doppler bias, giving the range estimate
\begin{equation}
R = \frac{c}{4k}\!\left( f_{b,\uparrow} - f_{b,\downarrow} \right).
\label{eq:range}
\end{equation}

\subsection{Altitude Estimation}
To extract altitude, the received signal $S_{R}(t)$ is processed using a conventional fast Fourier Transform (FFT)-based dechirping method. The upsweep and downsweep signal components are mixed with the known reference chirp to produce the beat signals, from which the beat frequencies $f_{b,\uparrow}$ and $f_{b,\downarrow}$ are extracted. Cell-averaging constant false alarm rate (CFAR) detection is applied to identify the ground return for each sweep, and the corresponding range estimate is obtained from (\ref{eq:range}). Averaging the upsweep and downsweep range estimates cancels residual Doppler bias and yields the altitude estimate, where the slant range to the ground surface directly beneath the aircraft is taken as the altitude.

In practice, interference, clutter, and noise degrade these measurements, corrupting the beat signal and introducing false detections or range estimation error. The TCAE described in the next section is applied to $S_{R}(t)$ prior to dechirping to suppress these unwanted components, producing a denoised reconstruction $S_{\mathrm{R,clean}}(t)$ that preserves the phase and frequency content necessary for accurate range estimation.

\subsection{IQ Signal Representation and Interference Model}
\label{se:iq_interference}

\subsubsection{IQ Representation and Model Input}
The received signal $S_R(t)$ in \eqref{eq:rx} is a real-valued bandpass 
signal centered at $f_c$. To obtain the complex baseband representation 
used by the TCAE, $S_R(t)$ is mixed with a complex exponential at the 
carrier frequency and low-pass filtered, yielding the complex envelope
\begin{equation}
    r(t) = \mathrm{LPF}\!\left\{ S_R(t)\, e^{-j2\pi f_c t} \right\},
    \label{eq:iq_demod}
\end{equation}
where $\mathrm{LPF}\{\cdot\}$ denotes low-pass filtering. The real and 
imaginary parts of $r(t)$ correspond to the in-phase (I) and quadrature 
(Q) components, respectively.

This continuous-time complex envelope is sampled at rate $F_s = B$, 
the radar signal bandwidth, to produce $N = 7500$ discrete-time IQ 
samples per chirp. The resulting sample vector is arranged as a 
two-channel real-valued input tensor
\begin{equation}
    \mathbf{x} = \bigl[\,\mathrm{Re}\{r[n]\},\; \mathrm{Im}\{r[n]\}\,\bigr]
    \in \mathbb{R}^{N \times 2},
    \label{eq:input_tensor}
\end{equation}
where $n = 0, 1, \ldots, N-1$. This is the input to the TCAE. The 
corresponding clean label used during training is
\begin{equation}
    \mathbf{y} = \bigl[\,\mathrm{Re}\{r_{\mathrm{clean}}[n]\},\; 
    \mathrm{Im}\{r_{\mathrm{clean}}[n]\}\,\bigr] \in \mathbb{R}^{N \times 2},
    \label{eq:label_tensor}
\end{equation}
where $r_{\mathrm{clean}}[n]$ is obtained from the same pipeline with 
$s_{\mathrm{intf}}(t) = s_{\mathrm{clutter}}(t) = n(t) = 0$. The TCAE 
is trained to map $\mathbf{x} \mapsto \mathbf{y}$ using MSE loss.

\subsubsection{Interference Model}
\label{se:interference_model}
Three classes of structured interference are considered, corresponding 
to realistic in-band threats in the 4.2--4.4 GHz band. In general, the interference term in \eqref{eq:rx} may include one or more of the following components, combined additively prior to sampling:
\begin{equation}
    s_{\mathrm{intf}}(t) = s_{\mathrm{tone}}(t) + s_{\mathrm{QPSK}}(t) 
    + s_{\mathrm{5G}}(t).
    \label{eq:intf_model}
\end{equation}
where any component may be set to zero. In the simulation and OTA experiments reported here, many evaluations consider a single interference type, but the model is trained on examples drawn from simultaneous interference sources.

\textit{Tone interference.} Continuous-wave (CW) tone interference 
consists of $K$ sinusoidal components at frequencies uniformly spaced 
within the radar bandwidth:
\begin{equation}
    s_{\mathrm{tone}}(t) = \sum_{k=1}^{K} A_k \cos(2\pi f_k t + \phi_k),
    \label{eq:tone}
\end{equation}
where $f_k \in [f_c - B/2,\, f_c + B/2]$, and the amplitudes $A_k$ and 
phases $\phi_k$ are randomized per training example.

\textit{QPSK burst interference.} Quadrature phase-shift keying (QPSK) 
interference is modeled as a short-duration burst occupying a randomized 
time interval $[t_0, t_0 + \Delta t]$ and frequency band within the 
radar chirp:
\begin{equation}
    s_{\mathrm{QPSK}}(t) = A_q\, p(t - t_0)\, \cos\bigl(2\pi f_q t 
    + \theta(t)\bigr),
    \label{eq:qpsk}
\end{equation}
where $p(t)$ is a rectangular pulse of duration $\Delta t$, $f_q$ is 
the burst carrier frequency, $\theta(t) \in \{0, \pi/2, \pi, 3\pi/2\}$ 
is the QPSK phase sequence, and $A_q$ is set to achieve a randomized 
signal-to-interference ratio (SIR) drawn uniformly from $[-20, 0]$ dB. 
Burst duration $\Delta t$ and bandwidth are also randomized per example.

\textit{5G downlink interference.} The 5G interference component 
$s_{\mathrm{5G}}(t)$ is synthesized using the MATLAB 5G New Radio (NR) 
waveform generator with 15 kHz subcarrier spacing and either 5 or 
10 MHz channel bandwidth. The waveform includes synchronization signals, 
control channels, and a full-band data region, and is aligned to 
partially overlap the radar chirp in both time and frequency.

\subsubsection{Interference Metrics and Temporal Overlap}
The SIR for a given example is defined as
\begin{equation}
    \mathrm{SIR} = 10\log_{10}\!\left( 
    \frac{\sum_{n=0}^{N-1} \lvert r_{\mathrm{clean}}[n] \rvert^2}
    {\sum_{n=0}^{N-1} \lvert s_{\mathrm{intf}}[n] \rvert^2} 
    \right) \;\mathrm{dB},
    \label{eq:sir}
\end{equation}
and the signal-to-interference-plus-noise ratio (SINR) additionally 
includes the additive noise contribution in the denominator. Both 
quantities are averaged over the $N$-sample input window.

Temporal overlap $\rho \in [0, 1]$ is defined as the fraction of the 
$N$-sample chirp window during which the interfering signal is active:
\begin{equation}
    \rho = \frac{\Delta t_{\mathrm{active}}}{N / F_s},
    \label{eq:overlap}
\end{equation}
where $\Delta t_{\mathrm{active}}$ is the duration of interference 
activity within the chirp window. A value of $\rho = 1$ indicates 
full temporal overlap.

\subsubsection{Doppler Modeling}
The Doppler frequency shift $f_D$ in \eqref{eq:rx} is given by the 
narrowband approximation
\begin{equation}
    f_D = \frac{2 v_r}{c} f_c,
    \label{eq:doppler}
\end{equation}
where $v_r$ is the radial velocity of the aircraft relative to the 
ground. In simulation, $v_r$ is derived at each time step from the 
loaded flight trajectory. As shown in \eqref{eq:beat}, $f_D$ appears 
as an additive bias on both upsweep and downsweep beat frequencies; 
averaging the two estimates in \eqref{eq:range} cancels this bias to 
first order. In the OTA experiments described in Section~\ref{se:ota}, 
no physical aircraft motion is present; altitude variation is emulated 
by synthetically shifting the beat frequency of the transmitted chirp, 
and Doppler effects are therefore not present in the hardware captures. 
This is acknowledged as a limitation of the current OTA evaluation.

\section{Temporal Convolutional Autoencoder Model and Training}

\subsection{Model Architecture}
The proposed model follows a standard autoencoder structure: the input is compressed to a latent representation and then expanded back to its original dimension. This bottleneck forces the network to generalize useful features while discarding nuisance components, enabling denoising. 

For RF applications, the network must handle complex-valued data. We adopt a two-channel representation, separating the real and imaginary parts of the IQ signal. This format allows standard 1D convolutions to mix and preserve frequency relationships. Alternative approaches, such as magnitude/phase encoding, were considered but found less effective for maintaining signal integrity. 

The encoder employs dilation factors of 1, 2, and 4, progressively expanding the temporal receptive field without increasing the number of convolutional kernel weights. The model incorporates dropout layers after each convolution to prevent overfitting, rectified linear unit (ReLU) activations for nonlinearity, and dilated convolutions to extend the receptive field without increasing depth. These elements allow the network to capture long-range temporal dependencies while remaining within a feedforward convolutional architecture. 

\begin{figure*}[h]
    \centering
    \includegraphics[width=0.95\textwidth]{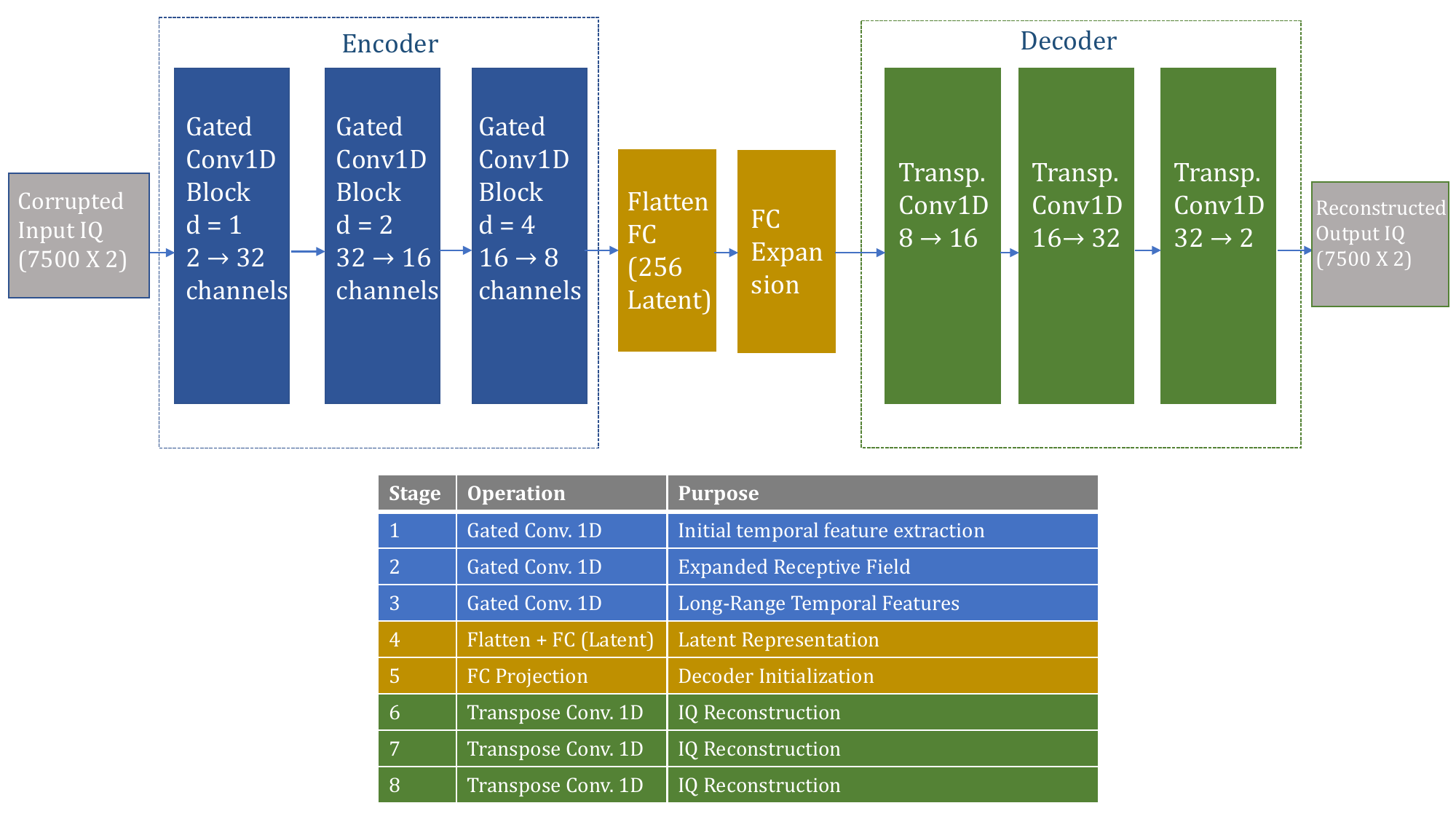}
    \caption{Architecture of the implemented TCAE. The encoder processes the corrupted two-channel IQ input using three gated one-dimensional convolutional blocks with progressively increasing dilation factors to capture temporal dependencies across the radar chirp. The resulting feature map is flattened and projected to a 256-dimensional latent representation, which is subsequently expanded through a dense projection and reconstructed by three transpose-convolutional blocks to produce the denoised IQ output. The table summarizes the principal stages of the implemented architecture. The exported implementation contains 17.29 million trainable parameters, with most parameters concentrated in the dense bottleneck projection layers (Table I).}
    \label{fig:TCN_Visual}
\end{figure*}

Building on our preliminary work with CNN and fully connected (FC) autoencoders~\cite{brown2024aircraft}, we introduce a TCAE architecture motivated by temporal feature extraction and practical model-integration constraints. Earlier FC-dominated models provided a useful proof of concept but incurred large parameter counts, while CNN-only variants gave promising denoising performance but presented implementation challenges in the Open Neural Network Exchange (ONNX)-based evaluation workflow used in this study. The TCAE was therefore selected as a practical architecture that combines dilated temporal convolutions with an encoder-decoder bottleneck and can be integrated into the downstream radar-altimeter processing chain. Figure~\ref{fig:TCN_Visual} summarizes the implemented TCAE architecture and the principal processing stages.

\subsection{Model Training}
The TCAE was trained on 50,000 FMCW radar signals, each containing 7500 IQ samples. An additional 10,000 examples were used for validation. The real and imaginary parts were normalized independently to unit peak amplitude before training. 

We trained the model using the Adam optimizer with a learning rate of $10^{-3}$, mean squared error (MSE) loss, and a batch size of 128 for 150 epochs on an NVIDIA GPU. The loss function directly compares reconstructed and clean signals, encouraging accurate recovery across the full signal dimension. 

Evaluation metrics included both RMSE between clean and reconstructed signals, and the accuracy of altitude estimates within the radar altimeter processing chain. To simulate realistic operating conditions, we augmented training with additive Gaussian noise, multipath fading via the simulation process described in Section~\ref{se:simstudy}, and multiple types of interference (tones, QPSK bursts, and 5G downlink waveforms). Hyperparameters such as kernel size, dilation factor, and compression ratio were tuned based on validation RMSE and range profile accuracy. The final model uses kernel sizes of 3–6, dilation factors up to 4, and a latent space representing 1.5–5\% of the input dimension.

\begin{figure*}[h]
    \centering
    \includegraphics[width=0.95\textwidth]{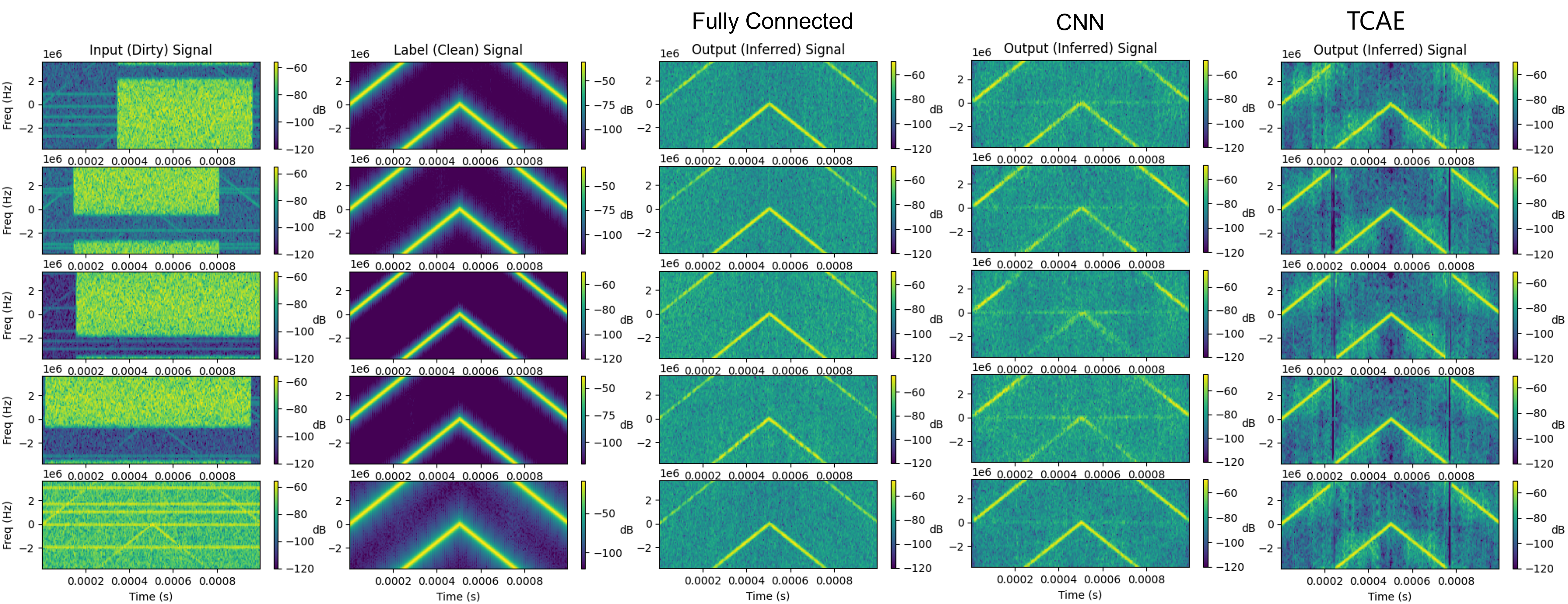}
    \caption{Under the same fixed training-data budget, the TCAE more consistently preserves the beat structure and suppresses residual interference artifacts than FC or CNN models, suggesting improved use of the available training examples in this setting.}
    \label{Model_Comparison}
\end{figure*}

Our preliminary work compared a CNN-only model with an FC model \cite{brown2024aircraft}. We observed that the CNN model outperformed the FC model in most scenarios and trained faster because of its reduced connectivity. Since then, we have expanded our evaluation to include a TCAE. For these comparisons, each model was trained on a dataset of 50,000 signals, which is lower than the amount of data typically used for optimal performance. The dataset contained signals with tone-based, quadrature phase-shift keying (QPSK), and 5G interference. We evaluated the models both quantitatively by measuring the RMSE between the predicted and clean signals and qualitatively by visually comparing the Short-Time Fourier Transform (STFT) of the output signals with those of the input and the training label. Although the FC and CNN models produced similar STFTs, there were differences in the RMSE of their output, with the CNN model performing better. The reconstructed signals showed visibly reduced interference artifacts in the STFT, along with improved beat signal clarity for range estimation. Figure \ref{Model_Comparison} shows the STFT results side by side, including the input signals and the clean evaluation labels, followed by the output of each model for 5 different evaluation examples.

The TCAE mixes the IQ components within its initial 1D convolution operations and leverages dilated convolutions to capture temporal correlations across the chirp. In the limited-data comparison shown in Fig.~\ref{Model_Comparison}, the TCAE better preserves the beat structure and suppresses residual interference artifacts relative to the FC and CNN-only variants trained under the same dataset size. We therefore interpret Fig.~\ref{Model_Comparison} as evidence of improved denoising performance under the fixed training budget used in this study, rather than as evidence of general parameter efficiency.

The primary distinction of the TCAE is its use of dilated 1D temporal convolutional blocks before and after a bottleneck projection. The dilation factors increase the temporal receptive field without increasing the number of convolutional kernel weights, which is useful for modeling long-range structure in FMCW IQ sequences. However, because the implemented model includes dense bottleneck projections, the total parameter count is dominated by these dense layers rather than by the convolutional blocks. Accordingly, we do not claim that the TCAE is lower in parameter count than all CNN-only autoencoder variants.

The implemented TCAE uses three gated 1D convolutional encoder blocks with channel dimensions $2 \rightarrow 32$, $32 \rightarrow 16$, and $16 \rightarrow 8$, respectively. Each block contains a feature convolution and a gate convolution with kernel size 3. The resulting feature map is flattened and projected to a 256-dimensional latent representation. The decoder first maps this latent vector through a dense projection and then reconstructs the two-channel IQ sequence using three transpose-convolutional layers with channel dimensions $8 \rightarrow 16$, $16 \rightarrow 32$, and $32 \rightarrow 2$. The exported encoder and decoder parameter counts are reported in Table~\ref{tab:tcae_param_audit}.

\begin{table}[t]
    \centering
    \caption{Trainable Parameter Count of the Exported TCAE}
    \label{tab:tcae_param_audit}
    \begin{tabular}{l r}
    \toprule
    \textbf{Component} & \textbf{Trainable Parameters} \\
    \midrule
    Encoder gated convolutional blocks & 4,336 \\
    Encoder dense bottleneck projection & 15,356,160 \\
    \midrule
    Total encoder & 15,360,496 \\
    \midrule
    Decoder dense projection & 1,926,472 \\
    Decoder transpose-convolutional blocks & 3,314 \\
    \midrule
    Total decoder & 1,929,786 \\
    \midrule
    Total exported TCAE & 17,290,282 \\
    \bottomrule
    \end{tabular}
\end{table}

The exported ONNX encoder and decoder were audited to determine the trainable parameter count shown in Table~\ref{tab:tcae_param_audit}. The implemented TCAE contains 17.29 million trainable parameters. Most parameters are concentrated in the dense bottleneck and decoder projection layers, while the convolutional layers account for only a small fraction of the total. Thus, the present architecture should not be interpreted as a minimal-parameter or embedded-optimized design. Rather, it is evaluated here as an effective IQ-domain interference-mitigation architecture that integrates with the simulation and OTA processing chain. Reducing the parameter count through strided temporal compression, fully convolutional bottlenecks, separable convolutions, pruning, quantization, or other architecture-optimization methods is left for future work.

The loss function employed for the TCAE compares the rebuilt inferred output signals to the clean label signals using MSE loss. The loss function compares the raw signals which allows for optimization over the full signal dimension.

\begin{figure*}
    \centering
    \includegraphics[scale=0.45]{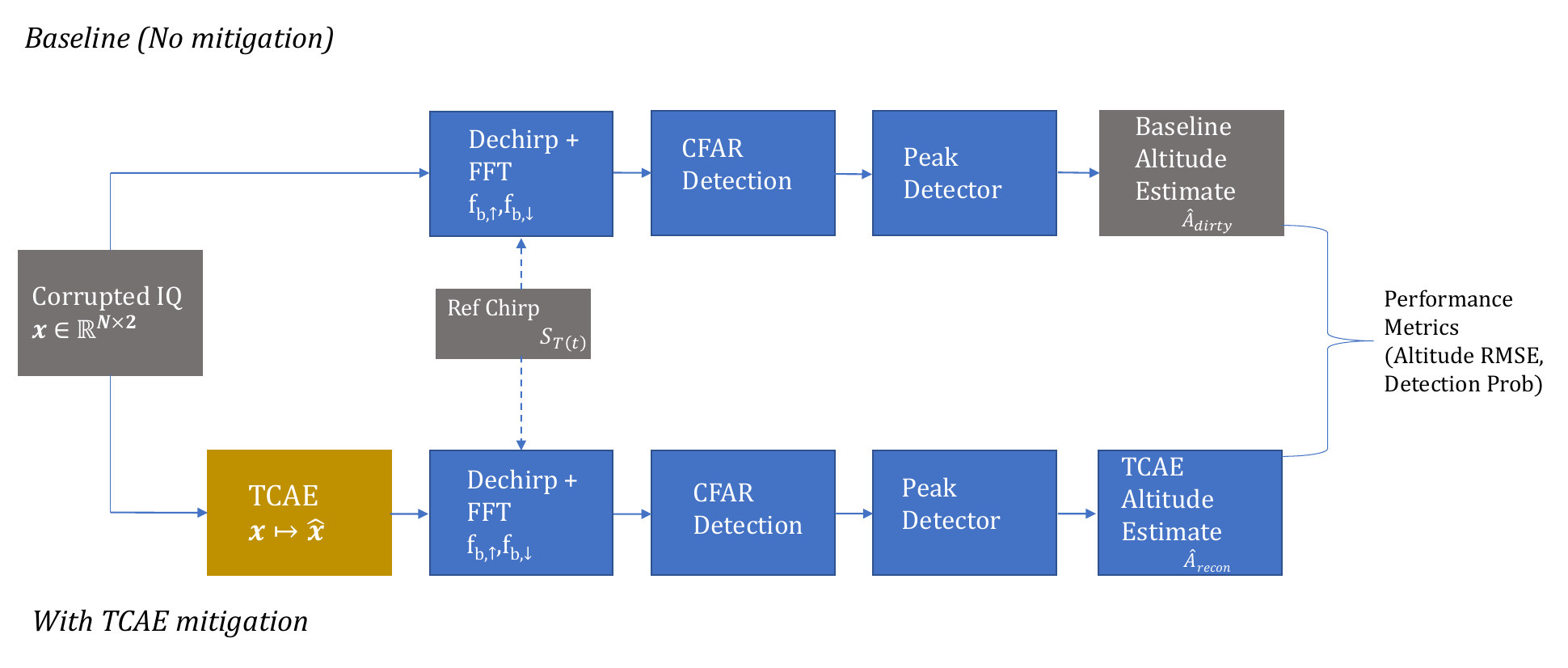}
    \caption{Range profile evaluation scheme used to assess TCAE performance. 
The corrupted IQ input $\mathbf{x}$ is processed along two parallel paths: 
a baseline path with no mitigation (top), and a TCAE-mitigated path 
(bottom). Both paths apply dechirping using the known reference chirp 
$s_T(t)$ to produce range profiles, from which altitude estimates are 
extracted via peak detection. Performance metrics are computed by 
comparing the estimated altitudes from each path against the true 
altitude derived from the flight trajectory.}
   \label{RangeProfileEvaluation}
\end{figure*}

\begin{figure*}
    \centering
    \includegraphics[scale=0.45]{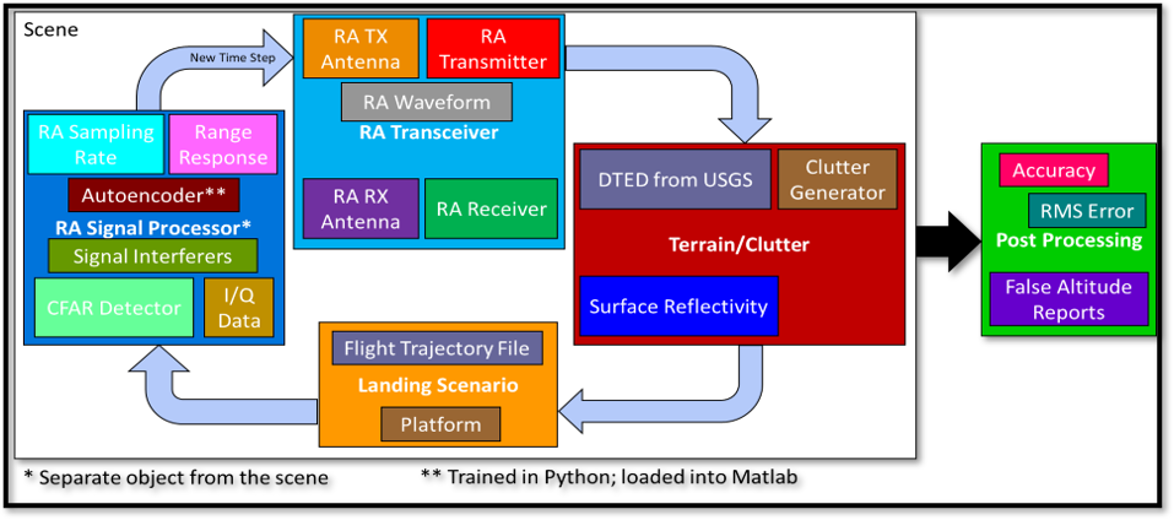}
    \caption{Block diagram of the FMCW radar altimeter simulation framework. The radar transmitter (RA TX) and receiver (RA RX) operate over a simulated landing scenario using real flight trajectory data and Digital Terrain Elevation Data (DTED). Clutter is generated from terrain and  surface reflectivity models. The TCAE (trained offline in Python and loaded into MATLAB via ONNX) is inserted into the RA Signal Processor prior to CFAR detection, operating on the received IQ data to suppress interference before range estimation.}
    \label{AltimeterSimulationDiagram}
\end{figure*}

\section{Simulation Study}
\label{se:simstudy}
\begin{figure*}
    \centering
    \includegraphics[scale=0.45]{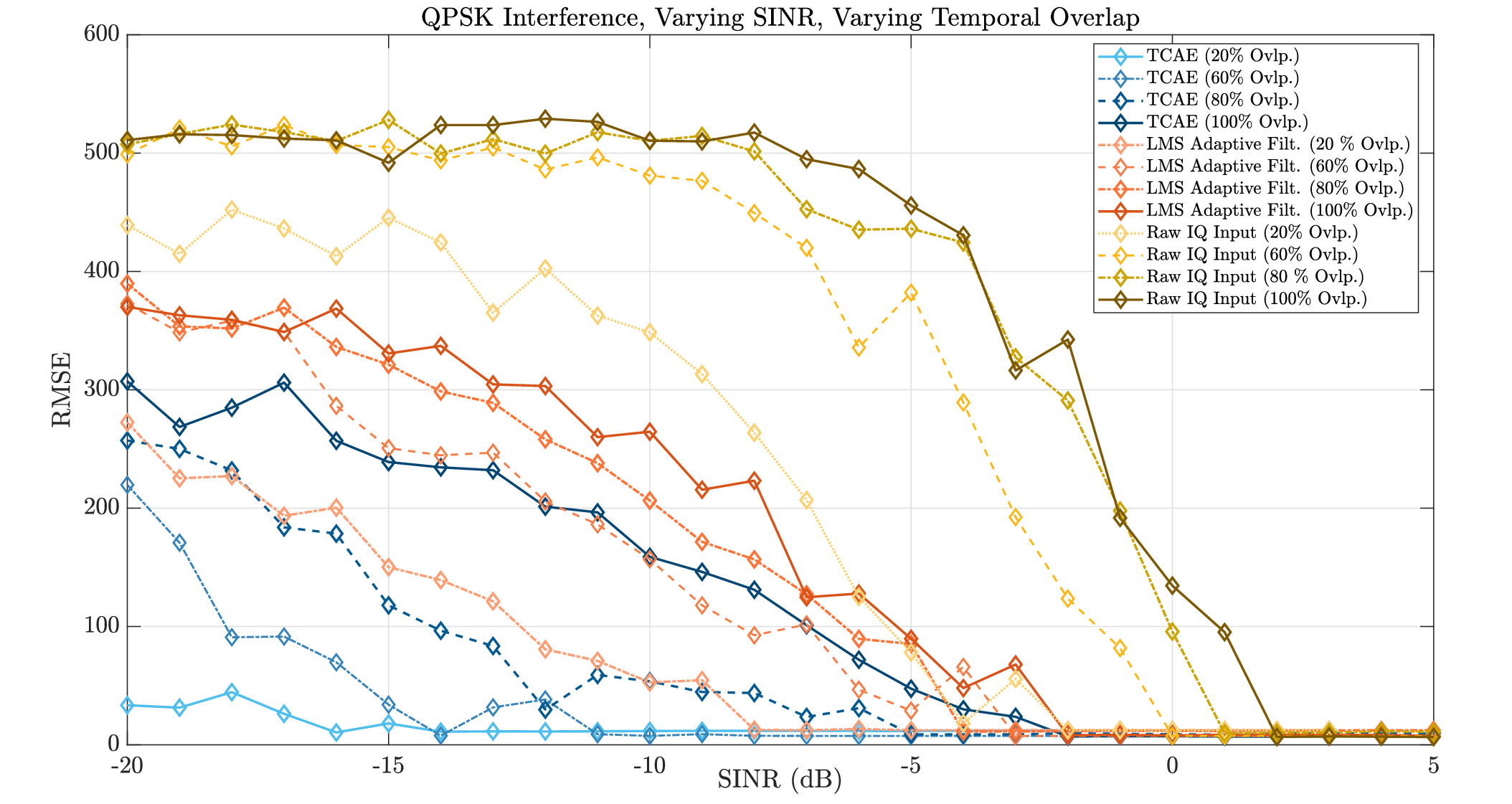}
    \caption{Range estimation accuracy under QPSK interference, measured by RMSE of altitude versus SINR for different temporal overlap levels. The raw input (no mitigation) degrades rapidly as overlap increases, and LMS adaptive filtering fails under high-overlap, low-SINR conditions. In contrast, the proposed TCAE maintains RMSE below 50 m even at 100\% overlap and –15 dB SINR, reducing error by more than 85\% relative to LMS. These results confirm the TCAE’s robustness to severe in-band interference.}
    \label{fig:bigCurve}
\end{figure*}

The TCAE model is evaluated within a comprehensive simulation of a commercial aircraft radar altimeter system based on the MATLAB FMCW Radar Altimeter Simulation framework \cite{Mathworks}. The simulation replicates a realistic landing scenario by integrating real-world data and carefully defined radar parameters as shown in Figure \ref{AltimeterSimulationDiagram}. Digital Terrain Elevation Data (DTED) from the United States Geological Survey (USGS) is imported from an actual elevation file, and a set of predefined reflectivity models is available to characterize various land types. In our simulation, a low-relief urban setting is selected to represent typical ground conditions for an airport environment. In addition, a real flight trajectory is loaded from a data file to capture an aircraft’s approach to Chicago O'Hare Airport. These data sources collectively define the land surface, over which clutter is generated to emulate the range response encountered by the radar altimeter during landing.

The radar altimeter simulation is built upon detailed parameter definitions generally based on International Telecommunication Union (ITU) recommendations. Key parameters include a center frequency of 4.3 GHz, a bandwidth of 7.5 MHz, and a frequency of 1000 chirps per second. The system models the altimeter antennas using a phased Gaussian element with a wide beamwidth (approximately 40 degrees) to ensure robust performance despite variations in aircraft attitude. The transmitted waveform is a triangular FMCW signal generated by calculating an appropriate sweep time and slope. To simulate realistic interference conditions, three types of interference were added to the FMCW radar returns: CW tones, QPSK-modulated bursts, and a 5G downlink waveform, as defined in Section~\ref{se:iq_interference}. Interference timing and power levels were randomized per example to encourage model generalization across SIR and temporal overlap $\rho$ conditions.

As a classical benchmark, we implemented a block least mean squares (LMS) adaptive filter using MATLAB's digital signal processing (DSP) toolbox to process the received signal prior to altitude estimation. The filter was configured with a length of 32, block size of 100, and step size $\mu = 10^{-4}$. The reference signal for adaptation was the clean transmitted chirp, and the filtered output was passed to the altimeter signal processor for range estimation. This baseline served to evaluate the extent to which traditional adaptive filtering could mitigate in-band interference. While the LMS filter achieved moderate performance under low interference overlap, it exhibited significant degradation when the interference duration or power increased, due to limited convergence and mismatch between the interference and reference signals. Compared to the LMS approach, the TCAE consistently achieved lower RMSE in reconstructed signals and more accurate altitude estimates, particularly under low-SIR or temporally overlapping interference conditions. For example, at 100$\%$ temporal overlap and –15 dB SINR, the TCAE reduced RMSE by over 85$\%$ relative to the LMS filter. The evaluation scheme used to compare these paths is illustrated in Fig.~\ref{RangeProfileEvaluation}, which shows how both the unmitigated and TCAE-processed IQ signals are passed through the same dechirping and detection chain to produce comparable altitude estimates.

Two sampling rates are employed: one for generating and dechirping the waveform (based on the signal bandwidth) and a second, lower rate for processing the dechirped beat signal. The radar transceiver component handles both transmission and reception, and its outputs are subsequently fed into a custom altimeter signal processor.

Within the signal processor, the dechirping process, in which the received signal is mixed with the known transmitted waveform, yields a beat frequency that is directly related to the target range. The processor converts these frequency measurements into range values using the known sweep slope and then applies a CFAR algorithm to detect the ground return from clutter. Detections from both the upsweep and downsweep signals are combined, and the effects of Doppler shift are minimized by averaging to produce a robust altitude estimate. The simulation continuously evaluates performance by comparing the estimated altitude with the true altitude derived from the flight trajectory and terrain data. Plots of range responses, detection results, and altitude error over time provide detailed insights into the altimeter’s performance throughout the landing scenario, where varying clutter characteristics and changing beam footprints as the aircraft descends are fully accounted for.

Figure \ref{fig:bigCurve} provides a quantitative comparison of range estimation accuracy across different interference mitigation strategies under varying SINR and interference overlap levels. The temporal overlap $\rho$ (defined in Section~\ref{se:iq_interference}) refers to the fraction of the chirp window during which the QPSK interferer is active. The RMSE is computed from the altimeter's estimated altitude over a simulated landing scenario. As shown, the raw input (no mitigation) exhibits significant performance degradation even under moderate interference, with RMSE exceeding 400~m at high overlap and low SINR. The LMS filter improves upon this baseline at partial overlaps but fails to converge effectively at full ($100\%$) interference overlap, particularly below –10 dB SINR. In contrast, the TCAE maintains robust performance across all overlap levels, achieving RMSE below 50~m even under severe conditions ($100\%$ overlap at –15 dB SINR). These results highlight the limitations of traditional adaptive filtering and demonstrate the TCAE's superior ability to learn and suppress structured in-band interference.

\begin{figure*}
    \centering
    \includegraphics[scale=0.45]{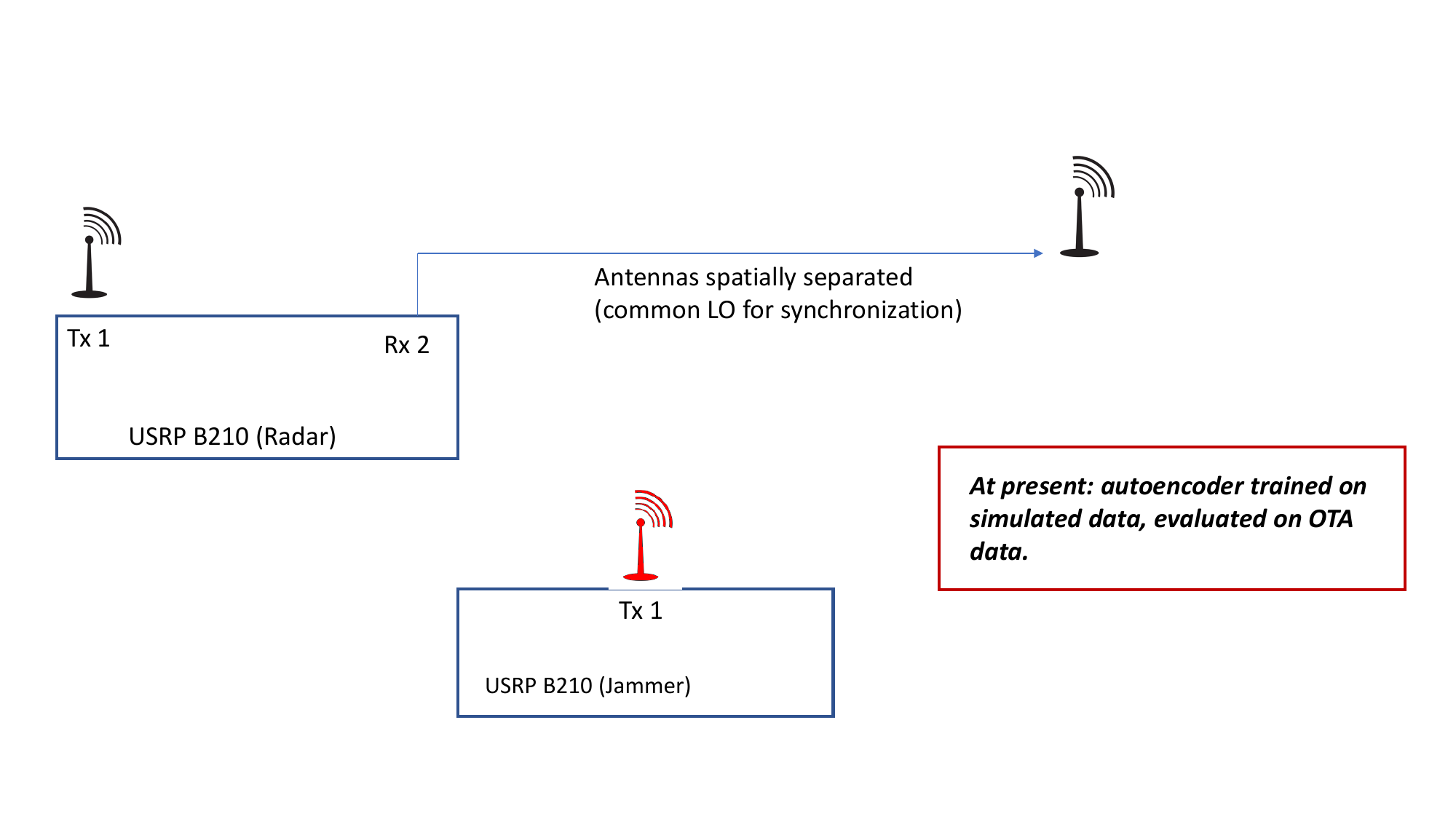}
    \caption{Over-the-air (OTA) testbed configuration. Two USRP B210 devices are used: the radar USRP transmits the FMCW chirp and receives the composite signal, while the jammer USRP transmits the interfering waveform. The autoencoder was trained entirely on simulated data and evaluated on these OTA captures.}
    \label{fig:ota}
\end{figure*}

\begin{figure*}
    \centering
    \includegraphics[scale=0.6]{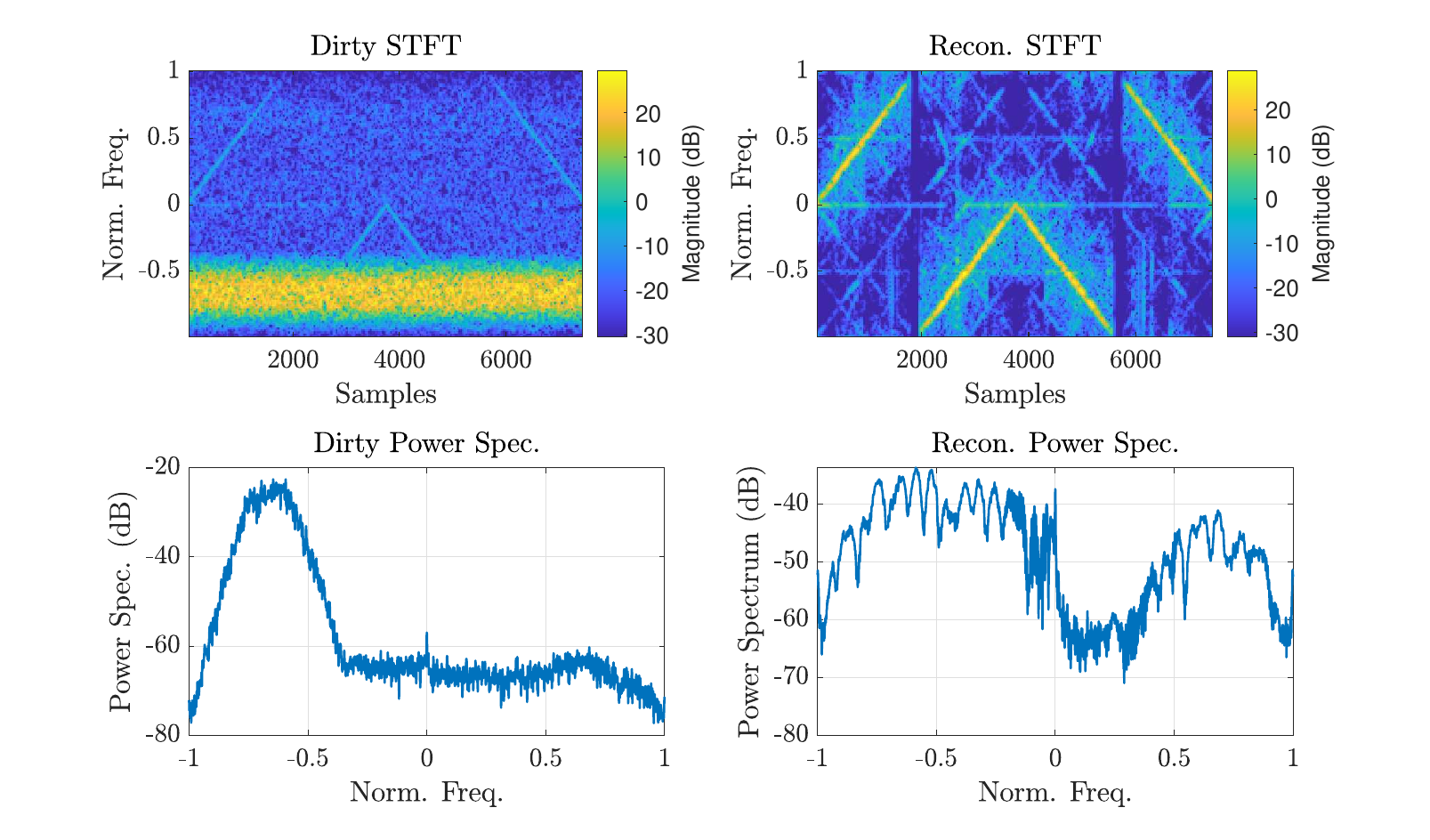}
    \caption{OTA evaluation of the TCAE. Left: STFT of the received FMCW signal corrupted by a high-power interferer occupying ~27\% of the bandwidth. Right: reconstructed output after TCAE processing. The autoencoder suppresses the interferer while preserving the beat structure of the radar return. Bottom: corresponding power spectra show that the TCAE removes the interferer without distorting the underlying FMCW chirp, demonstrating preservation of frequency content essential for altitude estimation.}
    \label{fig:jamRemove}
\end{figure*}

\begin{figure}
    \centering
    \includegraphics[scale=0.55]{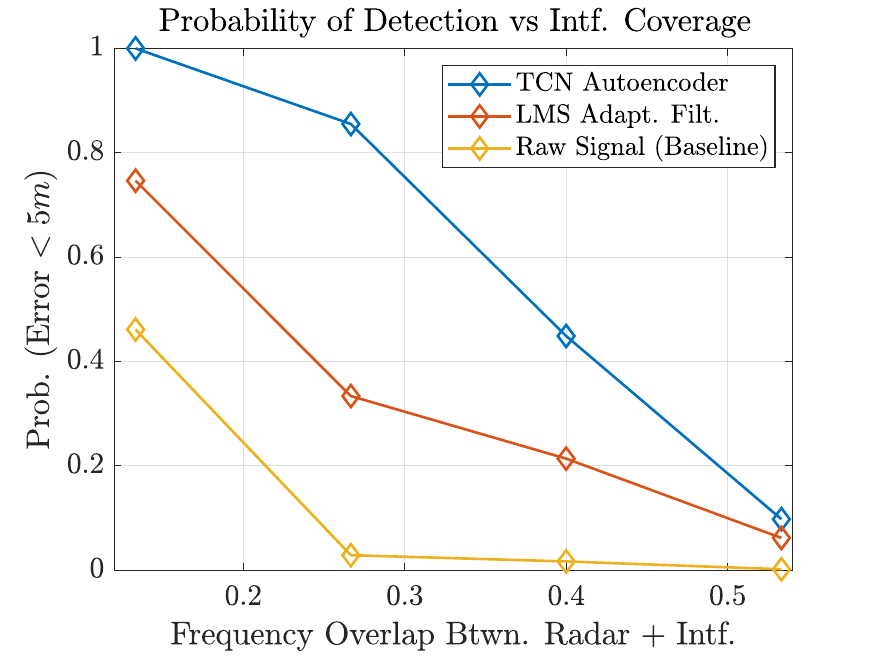}
    \caption{Probability of correct altitude detection (defined as error $<5$m) under OTA interference scenarios. The TCAE sustains high detection probability across a wide range of interference overlaps, whereas unmitigated input and LMS filtering degrade sharply beyond ~30\% overlap. This confirms that the TCAE generalizes from simulation to hardware tests and remains effective under laboratory conditions.}
    \label{fig:pd}
\end{figure}

These simulation results validate the effectiveness of the TCAE in mitigating a wide range of interference types under realistic altimeter conditions. Across various terrain types and flight paths, the TCAE model demonstrated consistent altitude estimation accuracy and robustness to tone, QPSK, and 5G-style interference. This confirms the model’s practical utility in operational altimeter scenarios.

\section{Results on Over-the-Air Data}
\label{se:ota}
The performance of the proposed TCAE was evaluated through both simulation and over-the-air (OTA) testing. In simulation, the model consistently reduced altitude estimation error across a wide range of interference scenarios. Compared to LMS adaptive filtering and unmitigated input, the TCAE achieved substantially lower RMSE, particularly under conditions of high overlap and low SINR. This demonstrates its ability to learn and suppress structured in-band interference that conventional methods cannot manage.

OTA testing further validated these findings. Using a USRP-based testbed, the autoencoder suppressed strong interferers occupying more than one-quarter of the FMCW bandwidth while preserving the beat structure of the radar return. Across multiple overlap conditions, the TCAE maintained high probability of correct altitude detection (defined as $< 5$ m error), confirming its ability to generalize from synthetic training data to real-world conditions. Together, the simulation and OTA results show that the TCAE provides robust interference mitigation for radar altimeters without sacrificing range accuracy.

After training and evaluating our models on MATLAB-generated data, we conducted a quasi real-time evaluation using over-the-air signals. A diagram of the OTA testbed configuration is shown in Figure \ref{fig:ota}. In this setup, MATLAB served as the primary driver script and called custom Python functions to interface with GNU Radio (an open-source software-defined radio framework), which controlled two USRP B210 devices. Our MATLAB script loads the autoencoder via the ONNX framework with a batch size of 1 and initializes the altimeter signal processor using the same parameters as in the data generation phase. Next, we launch the GNU Radio flowgraph, which is invoked from MATLAB through Python bindings. In the OTA testbed, two USRP transmitters were used: one to transmit the clean FMCW radar chirp and another to transmit a high-power interfering signal. Thus, the GNU Radio flowgraph includes two USRP sink blocks for transmission and one USRP source block for reception. For the evaluation shown in Figure \ref{fig:ota}, the interfering signal was a QPSK-modulated burst that temporally overlapped with the radar chirp and occupied approximately 26.7$\%$ of the radar bandwidth. The interference was designed to simulate a low-SIR challenge case, with interference power exceeding the radar return by up to 10 dB. The amount of temporal overlap between the radar chirp and interference was varied across evaluations to test model robustness. The overlap was defined as the fractional duration of the 7500-sample chirp during which the interfering signal was active. This scenario emulated a plausible airborne interference event, such as a communications signal from another nearby emitter operating in-band.

One sink transmits the clean FMCW signal of interest while the other transmits interference, and the USRP source captures the incoming signals, storing the IQ data in memory via a GNU Radio vector sink. The MATLAB script then initiates an altimeter landing scenario loop by shifting the signal of interest to simulate changes in beat frequency similar to a radar approaching its target. After clearing the IQ data vector, the script pauses for 0.02 seconds to allow data accumulation before importing the data into MATLAB. We applied cross-correlation with the known reference chirp to segment the received IQ data into individual frames for autoencoder processing. Finally, the autoencoder outputs are processed by the radar altimeter processor to estimate altitude from the denoised signals, and we evaluate the model's performance using RMSE across approximately 10 to 20 processed signals at each synthetic altitude.

The over-the-air tests further reinforce the performance observed in simulation. As shown in Figure \ref{fig:jamRemove}, the TCAE successfully removes a high-powered interference signal occupying over 25$\%$ of the FMCW bandwidth while preserving the underlying radar return. Figure \ref{fig:pd} shows a high probability of correct detection (defined by $<$5m error) across a range of interference overlaps, demonstrating robustness to challenging real-time conditions. These results indicate that the proposed approach can transfer from simulation to laboratory OTA captures and support further investigation under more realistic altimeter operating conditions.

\section{Conclusion and Future Work}
We have presented an effective TCAE for interference mitigation in FMCW radar altimeters. By operating directly on IQ data, this model suppressed in-band interference while preserving frequency content necessary for accurate range estimation. Extensive simulations demonstrated that the TCAE maintains altitude estimation accuracy under severe interference conditions, reducing RMSE by more than 85\% compared to LMS adaptive filtering at low SINR and full temporal overlap. 

Over-the-air experiments using a USRP-based testbed further validated these findings, showing that the TCAE generalizes effectively from synthetic training data to hardware scenarios and sustains a high probability of correct altitude detection across diverse interference overlaps.

The present results should be interpreted as a proof of concept for IQ-domain learned interference mitigation rather than as a final embedded implementation. The exported TCAE contains $17.29$ million trainable parameters, with most parameters concentrated in dense projection layers. Although the model performs inference through a single feed-forward pass and integrates with the ONNX-based evaluation chain, this work does not evaluate latency, memory footprint, power consumption, quantization, or certification readiness. These implementation questions remain important directions for future work.

Several limitations should be noted. First, the OTA evaluation emulates altitude variation by synthetically shifting the beat frequency and does not include physical aircraft motion or Doppler effects. Second, the current TCAE was selected for denoising performance and integration compatibility, not for minimal parameter count. Third, the evaluation demonstrates robustness under the tested interference, clutter, and OTA conditions, but does not constitute certification evidence for safety-critical avionics deployment.

Future work could extend this framework in several directions: 
\begin{enumerate}
    \item Testing across broader radar platforms and interference types.
    \item Investigating interpretable ML techniques to support reliability and certification concerns, which are essential in safety-critical avionics environments governed by standards such as DO-178C, the primary software certification standard for airborne systems.
    \item Exploring hybrid approaches which combine learned suppression with classical signal processing. 
\end{enumerate}
Such efforts could further improve resilience and pave the way for field deployment in operational aviation environments.

\bibliographystyle{IEEEtran}
%\IEEEtriggeratref{4}
\bibliography{aecBib.bib}{}

\end{document}